\documentclass{article}
\usepackage[pdftex]{color,graphicx}
\usepackage{amsmath,enumerate, amsfonts, ulem, url}
\usepackage{natbib}
\usepackage{algorithm, algorithmic}
\newcommand{\bm}[1]{\mbox{\boldmath{$#1 $}}}

\begin{document}
\baselineskip 18pt
\begin{center}
{\LARGE\textbf{Sparse varying-coefficient functional linear model}}
\end{center}
\begin{center}
{\large Hidetoshi Matsui}
\end{center}

\begin{center}
\begin{minipage}{14cm}
{
\begin{center}
{\it {\footnotesize 
Faculty of Data Science, Shiga University \\
1-1-1, Banba, Hikone, Shiga, 522-8522, Japan. \\
}}

\vspace{2mm}

{\small hmatsui@biwako.shiga-u.ac.jp}
\end{center}
\vspace{1mm} 

{\small {\bf Abstract:} 
	We consider the problem of variable selection in varying-coefficient functional linear models, where multiple predictors are functions and a response is a scalar and depends on an exogenous variable.  	
	The varying-coefficient functional linear model is estimated by the penalized maximum likelihood method with the sparsity-inducing penalty.
	Tuning parameters that controls the degree of the penalization are determined by a model selection criterion.  
	The proposed method can reveal which combination of functional predictors relates to the response, and furthermore how each predictor relates to the response by investigating coefficient surfaces.  	
	Simulation studies are provided to investigate the effectiveness of the proposed method.  
	We also apply it to the analysis of crop yield data to investigate which combination of environmental factors relates to the amount of a crop yield.  
}

\vspace{3mm}

}
\end{minipage}
\end{center}

\section{Introduction}
Functional data analysis (FDA) is one of the most useful techniques for analyzing longitudinal data sets by treating individual time-course observations as functions.  
The effectiveness of FDA is reported in various fields of applications such as medicine, biology, demography \citep{UlFi2013}, and several methodologies for FDA have attracted more and more attentions \citep{RaSi2005,HoKo2012, KoRe2017}.  

Many of the research on functional data analysis is based on multivariate analysis of functional data, and in particular, there are many works on regression analysis \citep{YaMuWa2005b,McHoSt_etal2014,Mo2015,ReGoSh_etal2017}.  
Functional linear model, one of the typical methodologies in FDA, considers a relationship between a predictor and a response, where either or both of them are given as functions.  
For example, when the predictor is a function whereas the response is a scalar, coefficient in the linear model is also given by a function, and we can investigate when and how the predictor relates to the response from it.  
The functional linear model with a scalar response can be easily applied to the framework of regression models for multivariate data such as logistic regression and generalized linear models \citep{Ja2002, MuSt2005, ArKoKa_etal2009a},.  
Several extensions of the functional linear models to the nonlinear framework are also reported, such as quadratic models \citep{YaMu2010} and additive models \citep{MuYa2008}, and they provides more flexible prediction performance.  
However, they are more difficult to interpret from the estimated model than the functional linear model.  
As an another extension to the functional linear model, \cite{CaSa2008} and \cite{WuFaMu2010} independently propose the varying-coefficient functional linear models (VCFLM), where the predictor is a function and the response is a scalar and it also depends on an exogenous variable.  

We consider the problem of variable selection in the functional regression model to investigate which combination of the functional predictors relates to the scalar response.  
One of the popular techniques for variable selection in ordinal linear models is to apply the regularization method with the sparsity-inducing penalties such as lasso \citep{Ti1996}, elastic net \citep{ZoHa2005} and so on.  
Methodologies about sparse regularization are comprehensively summarized in \cite{HaTiWa2015}.  
To achieve the variable selection in functional linear model, \cite{MaKo2011} consider applying the group lasso-type penalty \citep{YuLi2006} to the parameters derived by basis expansions.  
\cite{GeMaSt2013}, \cite{KaMaYa_etal2016}, \cite{MaUm2020} also approached the variable selection problem for functional regression models by the sparse regularization.  

The motivating data for our analysis comes from the data for crop yield of tomatoes.  
Long-term multistage tomatoes grow for about one year and bear fruit for most of the period.  
It is said that the amount of the yield strongly depends on several environmental factors such as temperature, solar radiation and CO$_2$ concentrations during the period from flowering to fruiting.
In addition, their relations may vary by season.
For example, the effect of the CO$_2$ concentration on the crop yield in summer and in winter may be different.   
However, this relation have not been explicitly clarified; which combination of environmental factors is important, and how this relation varies according to the season.  

To elucidate these relations, we propose a method for variable selection in VCFLM with multiple functional predictors by the sparse regularization.  
We call the resulting model the sparse varying-coefficient functional linear model (SVCFLM).   
The functional predictors are supposed to be expressed by truncated Karhunen-Lo\'eve expansion.  
Furthermore, we assume that the coefficient functions in the VCFLM are expressed by basis expansions, and then their coefficients, vectors rather than functions, are estimated by the penalized least squares method with an adaptive elastic net penalty \citep{ZoZh2009} with an idea of the standardized group lasso \citep{SiTi2012}.  
To derive the estimates of the coefficients, we apply the blockwise coordinate descent algorithm \citep{YuLi2006,RaLaLi_etal2009}.  
The choice of values of tuning parameters included in the model strongly relates to the result of the variable selection.  
To select appropriate values of them, we apply a model selection criterion.   
The effectiveness of the proposed method is investigated through some simulation studies in viewpoints of prediction and selection accuracies.  
We also report the application of the proposed method to the analysis of the crop yield data.  

This article is organized as follows.  
In Section 2, we review the VCFLM with multiple functional predictors and a scalar response.  
Section 3 introduces the method for estimating the parameters included in the VCFLM by the sparse regularization.  
Section 4 shows some simulation results, and then we report the result of applying the proposed method to the analysis of crop yield data in order to select the environmental factors that relate to the crop yield in Section 5.
Finally we summarize the main points in Section 6.  

\section{Varying-coefficient functional linear model}
Suppose that $p$ functional predictors $X_1(s), \ldots, X_p(s)$ relate to a scalar response $Y$, where the $Y$ also depends on observed time point $t$, and that we have $n$ sets of realizations of them denoted by $\{x_{i1}(s),\ldots, x_{ip}(s), y_i, t_i; i=1, \ldots, n\}$.  
Note that the response $y_i$ are centered so that they have zero mean and unit variance.  
Then we model the relationship between the predictors and the response as the following VCFLM:
\begin{align}
	y_{i} = \sum_{j=1}^{p} \int_{\mathcal S} x_{ij}(s) \beta_{j}\left(s, t_{i}\right) ds + 
	\varepsilon_{i}, 
	\label{eq:VCFLM}
\end{align}
where $\beta_j(s, t)$ is a coefficient function and $\varepsilon_1, \ldots, \varepsilon_n$ are errors that independently and identically follows a normal distribution with mean zero vector and variance $\sigma^2$.  

We assume that the functional predictor $x_{ij}(s)$ are expressed by the truncated Karhunen-Lo\'eve expansion 
\begin{align}
		x_{ij}(s) = \sum_{k=1}^{m_1}\xi_{ijk}\phi_{jk}(s) = \bm\xi_{ij}^T \phi_{j}(s),
\end{align}
where $\bm{\phi}_j(s) = (\phi_{j1}(s), \ldots, \phi_{jm_1}(s))^T$ is a vector of eigenfunctions of the covariance function $c_j(s, t) = \frac{1}{n}\sum_{i=1}^{n}(x_{ij}(s)-\overline x_j(s))(x_{ij}(t)-\overline x_j(t))$ with $\overline x_j(t) = \frac{1}{n}\sum_{i=1}^n x_{ij}(t)$, and $\bm{\xi}_{ij} = (\xi_{ij1},\ldots, \xi_{ijm_1})^T$ is a vector of functional principal component (FPC) scores which is given by $\xi_{ijk} = \int_{\mathcal S}\left(x_{ij}(s) - \overline{x}_j(s)\right)\phi_{jk}(s)ds$.  
In addition we assume that the coefficient function $\beta_{j}(s, t)$ is expressed by basis expansions as follows:
\begin{align}
	\beta_{j}(s, t)=\sum_{k=1}^{m_1}\sum_{l=1}^{m_2} \phi_{jk}(s)b_{jkl}\psi_{l}(t)
	= \bm\phi_{j}(s)^T B_{j}\bm\psi(t),
	\label{eq:beta}
\end{align}
where $\bm\psi(t) = (\psi_1(t),\ldots, \psi_{m_2}(t))^T$ is a vector of basis functions and $B_j=(b_{jkl})_{kl}$ is a matrix of parameters.

Then the VCFLM (\ref{eq:VCFLM}) is expressed as follows.
\begin{align}
	y_{i} = 
	\sum_{j=1}^{p}\left\{\bm\psi(t)^T \otimes \bm w_{ij}^T\right\} \operatorname{vec} B_{j} + \varepsilon_{i},
\end{align}
where vec is a column-wise vectorization operator of a matrix.  
Finally it is expressed in the same form as the traditional linear model
\begin{align}
	\bm y = \sum_{j=1}^{p}Z_j\bm b_j + \bm\varepsilon,  
	\label{eq:VCFLM2}
\end{align}
where 
\begin{align*}
	Z_j &=
	\left(\bm\psi(t) \otimes \bm w_{1j}, \ldots, \bm\psi(t) \otimes \bm w_{nj}\right)^T,~~
	\bm b_j = \operatorname{vec} B_j, \\
	\bm y &= \left(y_1,\ldots, y_n\right)^T, ~~
	\bm\varepsilon = \left(\varepsilon_1,\ldots, \varepsilon_n\right)^T.
\end{align*}
Therefore, under the assumptions of the basis expansions, the problem of estimating the VCFLM (\ref{eq:VCFLM}) corresponds to that of estimating vectors of parameters $\bm{b}_1,\ldots, \bm{b}_p$ in (\ref{eq:VCFLM2}).
\section{Estimation}
We estimate the unknown parameter $\beta_j(s,t)$ $(j=1,\ldots, p)$ in the VCFLM (\ref{eq:VCFLM}) by the penalized least squares method with the sparsity inducing penalty.  
The penalized least squares criterion is given by
\begin{align}
	\label{eq:penls1}
	\frac{1}{n}\sum_{i=1}^{n}\left\{y_{i}\left(t_{i}\right) - \sum_{j=1}^{p} \int x_{ij}(s) \beta_{j}\left(s, x_{i}\right) ds \right\}^2 + \sum_{j=1}^{p}P_\lambda(\beta_j),
\end{align}
where $P_\lambda(\beta_j)$ is a penalty function, and here we apply a group adaptive elastic net penalty:
\begin{align}
	P_\lambda(\beta_j) = 
	\alpha\lambda\hat w_j\left[\sum_{i=1}^n\left\{\int x_{ij}(s) \beta_{j}\left(s, x_{i}\right)ds\right\}^2\right]^{1/2} + 
	\frac{1}{2}(1-\alpha)\lambda \sum_{i=1}^n\left\{\int x_{ij}(s) \beta_{j}\left(s, x_{i}\right)ds\right\}^2,
	\label{eq:pen}
\end{align}
where $\|\cdot\|_2$ denotes an $L_2$ norm, $\lambda\in [0, \infty)$ is a regularization parameter that controls the degree of penalty and $\alpha\in [0,1]$ is a tuning parameter of the elastic net. 
Furthermore, $\hat w_j$ is an adaptive weight to originally ensure the oracle property for coefficients of ordinal linear models \citep{ZoZh2009}.  
In addition, we penalize on $\int x_{ij}(s) \beta_{j}\left(s, t\right)dt$, not on the coefficient function $\beta_j(s,t)$, using an idea of the standardized group lasso \citep{SiTi2012}.  
There are two reasons for applying the standardized group lasso rather than ordinal group lasso \citep{YuLi2006}. 
First, it prevents differences in the degree of penalty on the coefficients due to the differences in scale among predictors.  
Second, by using it we can apply the blockwise coordinate descent algorithm described later, whereas it is difficult for the ordinal group lasso.  
The minimization problem (\ref{eq:penls1}) with the penalty (\ref{eq:pen}) is  transformed into 
\begin{align}
	\frac{1}{n}\left\|\bm y - \sum_{j=1}^{p}Z_j\bm b_j\right\|_2^2 + 
	\alpha\lambda\sum_{j=1}^p\hat w_j\|Z_j\bm{b}_j\|_2 + 
	\frac{1}{2}(1-\alpha)\lambda \sum_{j=1}^p\|Z_j\bm{b}_j\|_2^2.
	\label{eq:penls2}
\end{align}
Here the adaptive weight $\hat w_j$ is given by $\hat w_j = \|Z_j\tilde{\bm{b}}_j\|_2^{-1}$ with a least squares estimator $\tilde{\bm{b}}_j$.  

To derive the estimator of $\bm{b}_j$, we apply the blockwise coordinate descent algorithm for minimizing (\ref{eq:penls2}).
Before that, we apply the QR decomposition to the matrices $Z_j$ $(j=1,\ldots, p)$ so that $Z_j = U_jR_j$, where $U_j$ is an orthogonal matrix that satisfies $U_j^TU_j = nI$ and $R_j$ is an  upper triangular matrices.  
Denote $\bm{\theta}_j = R_j\bm{b}_j$, then the penalized least squares criterion (\ref{eq:penls2}) is expressed as
\begin{align}
	\frac{1}{n}\left\|\bm y - \sum_{j=1}^{p}U_j\bm\theta_j\right\|_2^2 + 
	\alpha\lambda\sum_{j=1}^p\hat w_j\|\bm\theta_j\|_2 + 
	\frac{1}{2}(1-\alpha)\lambda \sum_{j=1}^p\|\bm\theta_j\|_2^2.
	\label{eq:penls3}
\end{align}
Given the parameter vector for variables other than $j$-th variable, $\bm{\theta}_k$ $(k\neq j)$,  then $\bm{\theta}_j$ is updated as
\begin{align}
	\hat{\bm{\theta}}_j = \frac{1}{n(1-\alpha)\lambda + n}\mathcal{S}\left(U_j^T\bm r_j, n\alpha\lambda \hat w_j\right),
	\label{eq:thetahat}
\end{align}
where $\mathcal{S}(\bm x, \kappa) = (1-\frac{\kappa}{\|\bm x\|_2})_+\bm x$ with an arbitrary vector $\bm x$ and a scalar $\kappa$ and $(a)_+ = \max\{a, 0\}$.  
Furthermore, $\bm r_j = \bm y - \sum_{k\neq j} U_k\bm{\theta}_j$ are partial residuals.  
We can apply the blockwise coordinate descent algorithm to the minimization of (\ref{eq:penls3}) since the matrices $U_j$ are orthogonal, whereas it is difficult to apply it to minimize (\ref{eq:penls2}) directly, since the matrices $Z_j$ are not orthogonal.  

The blockwise coordinate descent algorithm is summarized as follows.
\begin{enumerate}
	\item Apply the QR decomposition to $Z_j$ for $j=1, \ldots, p$ and then calculate $U_j$ and $R_j$.  
	\item Set initial values for $\bm\theta_1, \ldots, \bm{\theta}_p$.  
	\item Calculate $\bm r_j$ using updated $\bm\theta_j$.
	\item For $j=1, \ldots, p$, update $\bm{\theta}_j$ by (\ref{eq:thetahat}).  
	\item repeat Steps 3 and 4 until convergence.  
\end{enumerate} 
Finally we obtain the estimator of $\bm b_j = {\rm vec}B_j$ as $\hat{\bm{b}}_j = R_j^{-1}\hat{\bm{\theta}}_j$ and estimated coefficient surface as $\hat{\beta}_j (s,t) = \bm{\phi}_j^T(s)\hat B_j\bm{\psi}(t)$.  

The estimated model strongly depends on the values of tuning parameters including $\lambda$, $\alpha$, and number $m_2$ of basis functions $\bm{\psi}(t)$, and in particular the choice of these values directly affects the result of the variable selection.  
Therefore, the choice of the tuning parameters is an crucial issue. 
To do it, we use a BIC-type model selection criterion, which is given in the form of
\begin{align}
	BIC = -n\log\hat{\sigma}^2 + \widehat{df}\log n,
	\label{eq:BIC}
\end{align}
where $\widehat{df}$ is an effective degrees of freedom of the statistical model.  
\cite{RaLaLi_etal2009} derived an effective degrees of freedom of the sparse additive models, and \cite{WaLiTs2007} and \cite{ZhLiTs2010} also derive similar effective degrees of freedom of the model estimated by the sparse regularization.  
Applying these results to our method, the effective degrees of freedom of SVCFLM is given by
\begin{gather*}
	\widehat{df} = \sum_{j=1}^p {\rm tr}\left\{(I+\Omega_j)^{-1}\right\}I(Z_j\hat{\bm b}_j=\bm 0),\\
	\Omega_j = {\rm diag}\left\{\alpha\lambda\hat w_j \frac{1}{\|\hat{\bm b}_j\|_2} + 
	(1-\alpha)\lambda\right\}.
\end{gather*}
We select the tuning parameters that minimize the BIC and then treat the corresponding model as an optimal one.  
\section{Simulation study}
We conducted Monte Carlo simulations to evaluate the effectiveness of the proposed method. 
In this simulation we artificially generated the data for functional predictors, a scalar response and an exogenous variable, and then investigated whether the proposed method gives accurate results in viewpoints of prediction and variable selection performance.  

The setting of the simulation study is in the following.  
First, we generated the functional predictors $X_j(s)$ and the response $Y$ according to the VCFLM (\ref{eq:VCFLM}), where $X_j(s)$ are supposed to be expressed by $B$-spline basis $\bm\varphi_j(s)$ with random coefficients $\bm w_{ij}$, that is, $i$-th observation is given by $g_{ij}(s) = \bm w_{ij}^T\bm{\varphi}_j(s)$.  
In fact, the data for $X_j(s)$ are observed at discrete time points rather than continuously, and therefore the longitudinal data are generated at discrete time points with additional Gaussian noises $\epsilon_i$ with mean zero and standard deviation $\tau = 0.1R_{ij}^X$ with $R_{ij}^X = \max_\alpha g_{ij}(s_\alpha) - \min_\alpha g_{ij}(s_\alpha)$.  
That is, the longitudinal data corresponding to the predictors, $x_{ij\alpha}$ $(\alpha=1,\ldots, N)$, are given by
\begin{align*}
	x_{ij\alpha} = g_{ij}(s_\alpha) + \epsilon_i.  
\end{align*}
The number of time points is set to be $N=21$ for all subject.  
Coefficient functions $\beta_j(s,t)$ $(j=1,\ldots, p)$ are also supposed to be expressed by basis expansions like (\ref{eq:beta}), where the elements of the matrices $B_j$ for $j=1, \ldots, p/2$ are generated from normally distributed random values and $B_j = O$ for $j=p/2+1, \ldots, p$ so that the half of the predictors are irrelevant to the response.   
Then we generated the response by the following VCFLM:
\begin{align*}
	y_{i} = f(x_{i1}, \ldots, x_{ip}) + 
	\varepsilon_{i},~~
	f(x_{i1}, \ldots, x_{ip}) = \sum_{j=1}^{p} \int_{\mathcal S} x_{ij}(s) \beta_{j}\left(s, t_{i}\right) ds.
\end{align*}
We assume in this simulation that the errors $\varepsilon_1, \ldots, \varepsilon_n$ in the VCFLM are i.i.d and that their standard deviation is $\sigma = sR^Y$ with $s=0.1, 0.3$ and $R_Y = \max_i f(x_{i1}, \ldots, x_{ip}) - \min_i f(x_{i1}, \ldots, x_{ip})$.  
The purpose of this simulation is to evaluate whether the proposed method appropriately predicts the response and selects the functional predictors.  

We then apply the proposed SVCFLM to the generated data.  
First, we transformed the generated longitudinal data $x_{ij1}, \ldots, x_{ijN}$ into functional data $x_{ij}(s)$ for all $i$ and $j$ using KL expansions via basis expansions, and then derive FPC scores $\bm\xi_{ij}.$  
Here we used $B$-splines for the basis functions $\bm{\psi}$.  
We used an R package {\tt fda} for transforming the longitudinal data into functional data and for obtaining FPC scores.  
Then we estimated the VCFLM by minimizing the penalized least squares criterion (\ref{eq:penls1}) with the adaptive group elastic net penalty.  
Here we fixed the number $m_2$ of basis functions $\bm{\psi}(t)$ and the tuning parameter $\alpha$, and selected the regularization parameter $\lambda$ by BIC in (\ref{eq:BIC})

To evaluate the prediction accuracy of the proposed method, we calculated the root mean squared error (RMSE) given by
\begin{align*}
	{\rm RMSE} = \left[\frac{1}{n}\sum_{i=1}^{n}\left\{
	f(x_{i1}, \ldots, x_{ip}) - \widehat f(x_{i1}, \ldots, x_{ip})
	\right\}^2\right]^{1/2},  
\end{align*}
where $\widehat f(x_{i1}, \ldots, x_{ip})$ is an estimated function of $f(x_{i1}, \ldots, x_{ip})$.  
In addition to the RMSE, we also calculated accurate positive rates (APR) and accurate negative rates (ANR) about the variable selection, which are respectively given as follows.
\begin{align*}
	{\rm APR} = \frac{\#\{j|\widehat\beta_j\neq 0, \beta_j\neq 0\}}{\#\{j|\beta_j\neq 0\}}, ~~
	{\rm ANR} = \frac{\#\{j|\widehat\beta_j = 0, \beta_j= 0\}}{\#\{j|\beta_j= 0\}},
\end{align*}
where $\#\{j|A_j\}$ means the number of $j\in\{1,\ldots, p\}$ that satisfies an event $A_j$.  
We compared these prediction and variable selection accuracy of the proposed method with those of the ordinal functional linear models with sparse estimation (SFLM).  
We also compared the results of the estimation with and without the adaptive weight.  
We repeated this process 100 times, and then calculated the averaged values of 100 RMSEs and their standard deviations, APRs and ANRs.  

\begin{table}[t]
	\centering
	\caption{Results for simulation studies.  aSVFLM and aSFLM are SVFLM and SFLM with adaptive weights respectively.  }
	\label{tab:sim}
	\begin{tabular}{ccccc}
		\hline
		& SVCFLM & aSVCFLM & SFLM & aSFLM \\ 
		\hline 
		\multicolumn{5}{c}{$n=100, s=0.1$}\\
		\hline
		RMSE & 7.23$\times 10^{-1}$ & 2.76$\times 10^{-1}$ & 9.16$\times 10^{-1}$ & 8.94$\times 10^{-1}$ \\ 
		APR(\%) & 81.80 & 94.80 & 91.20 & 91.00 \\ 
		ANR(\%) & 86.40 & 85.80 & 27.00 & 30.40 \\ 
		\hline
		\multicolumn{5}{c}{$n=100, s=0.3$}\\
		\hline
		RMSE & 5.91$\times 10^{-1}$ & 2.98$\times 10^{-1}$ & 7.46$\times 10^{-1}$ & 7.42$\times 10^{-1}$ \\ 
		APR(\%) & 70.00 & 94.00 & 85.00 & 81.60 \\ 
		ANR(\%) & 87.80 & 87.00 & 17.20 & 21.60 \\ 
		\hline
		\multicolumn{5}{c}{$n=200, s=0.1$}\\
		\hline
		RMSE & 1.93$\times 10^{-1}$ & 0.51$\times 10^{-1}$ & 8.70$\times 10^{-1}$ & 8.70$\times 10^{-1}$ \\ 
		APR(\%) & 99.20 & 100.00 & 88.60 & 82.20 \\ 
		ANR(\%) & 30.00 & 63.60 & 13.20 & 25.00 \\
		\hline
		\multicolumn{5}{c}{$n=200, s=0.3$}\\
		\hline
		RMSE & 2.06$\times 10^{-1}$ & 1.19$\times 10^{-1}$ & 8.76$\times 10^{-1}$ & 8.78$\times 10^{-1}$ \\ 
		APR(\%) & 99.20 & 100.00 & 90.80 & 82.00 \\ 
		ANR(\%) & 25.60 & 66.00 & 8.80 & 22.20 \\ 
		\hline
	\end{tabular}
\end{table}

Table \ref{tab:sim} shows the results of the simulation studies.  
For all settings the SVCFLM gives smaller RMSEs than SFLM, and in particular, SVCFLM with an adaptive weight significantly reduces the RMSE.  
On the other hand for SFLM, there is no significant difference in RMSEs whether or not the adaptive weights are given.  
In addition, APRs and ANRs for SVCFLM with an adaptive weight are higher than other method.  
In particular, results for SFLM and aSFLM gives smaller ANRs, which indicates that they tend to select more variables than needed.  
These result shows that the SVCFLM can appropriately predict the response and select correct variables, especially with the adaptive weight.  

\section{Analysis of crop yield data}
\begin{figure}[t]
	\begin{center}
		\includegraphics[width=1.0\hsize]{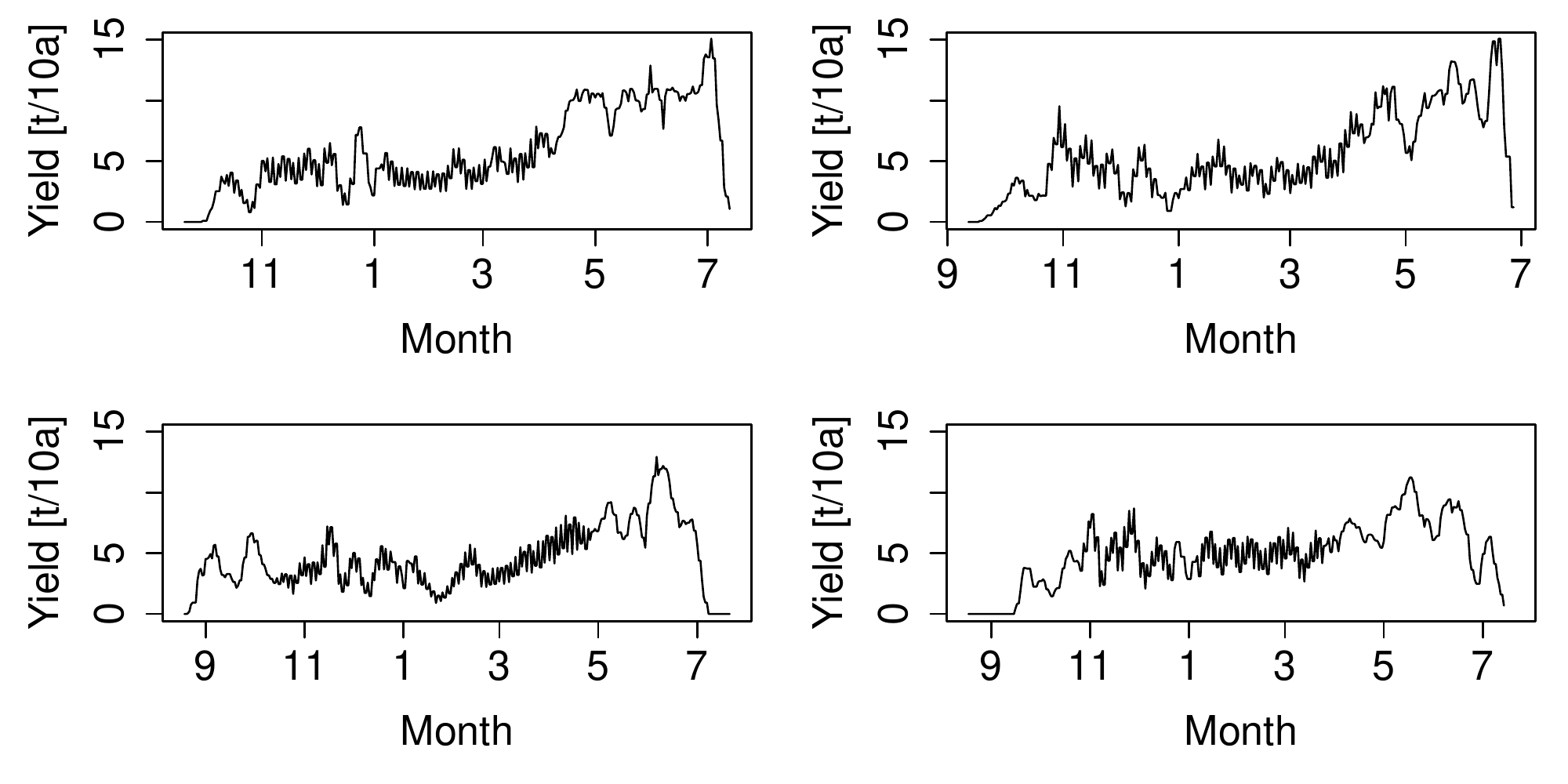}
	\end{center}
	\caption{Amount of crop yield for four terms; 2017 to 2020 from top left to bottom right.  The moving averages for 7 days are shown for all figures. } 
	\label{fig:tomato_data}
\end{figure}
%

We report the result of the application of the proposed method to the analysis of crop yield data for multi-stage tomatoes cultivated in a greenhouse on a farm in Kobe, Japan.  
Each seedling of the multi-stage tomatoes grows for about one year from August to next July, and are harvested almost every day from October to next July. 
In this study, we used daily yield data of single breed measured from October 2017 to July 2021.  
We treat the harvest period from August to next July as one term, then the data contains four terms (from 2017 to 2020).  
In addition to them, several kinds of environmental factors such as temperature and CO$_2$ concentration are repeatedly measured every minute by measuring equipment installed inside and outside the greenhouse.  
The objective of this analysis is to select environmental factors that relate to the amount of crop yields via regression modeling, treating the environmental factors as functional predictors and the daily crop yields as a scalar response.  

The original data for the yield vary greatly because the yield is zero when the farm is on a holiday, which makes the analysis difficult.  
To deal with this problem, we used the moving averages of the yield up to seven days from the harvest date.  
Figure \ref{fig:tomato_data} shows the data for tomato yields for four terms, where the data are moving averaged up to seven days.  
On the other hand, we used 39 longitudinally observed environmental factors that are considered to relate to the crop yield, averaged over 1 hour, as predictors.    

It is considered that the growth of the tomato fruits are influenced by the environmental factors during about 80-day period before fruiting.  
We used hourly averaged data for the environmental factors as functional data, and then constructed a regression model, treating these environmental factors corresponding to 80 days before the maturing day as functional predictors and the daily yield of tomatoes as a response.   
In this analysis we treat a set of an yield of certain day and 80-day environmental factors before the day corresponds to an individual.  
For example, the $i$-th observation includes the amount of crop yield $y_i$ at day $t_i$ and the environmental factors $x_{ij}(t)$ $(j=1,\ldots, 39)$ for 80 days before day $t_i$.  
The $(i+1)$-th observation corresponds to those on the next day.  
By merging the data of four terms, we have a sample with size $n=1093$.  
Furthermore, it is also considered that the influence of the environmental factors on the crop yields differs by season of the years, which is what motivated us to use the VCFLM, and therefore we treat the day $t_i$ of the year as an exogenous variable.  
Note that, since in this analysis the number of time points for the response is one for each individual and the observed period of the dataset is only four years, it is difficult to apply the function-on-function regression model \citep{RaDa1991}.  

We transformed the data for the environmental factors into functional data via basis expansion and then calculated the FPCs by Karhunen-Lo\'eve expansions.  
For such values we then applied the VCFLM (\ref{eq:VCFLM}), where $s$ and $t$ correspond to the day before cultivation and the day of the year of the cultivation, respectively.  
The model is estimated by the penalized least squares criterion with the adaptive elastic net regularization, given in (\ref{eq:penls3}), and the tuning parameters are selected by BIC.  

The sparse VCFLM selects 22 variables out of 39 variables. 
The selected environmental factors are listed in Table \ref{tab:env}.  
It is said from the farmer that the temperature, CO$_2$ concentration and solar radiation significantly relate to the crop yield, and our method could select these factors.  
On the other hand, derived factors such as cumulative solar radiation and temperature difference between day and night are not selected in the model.  

\begin{figure}[t]
	\begin{center}
		\includegraphics[width=0.48\hsize]{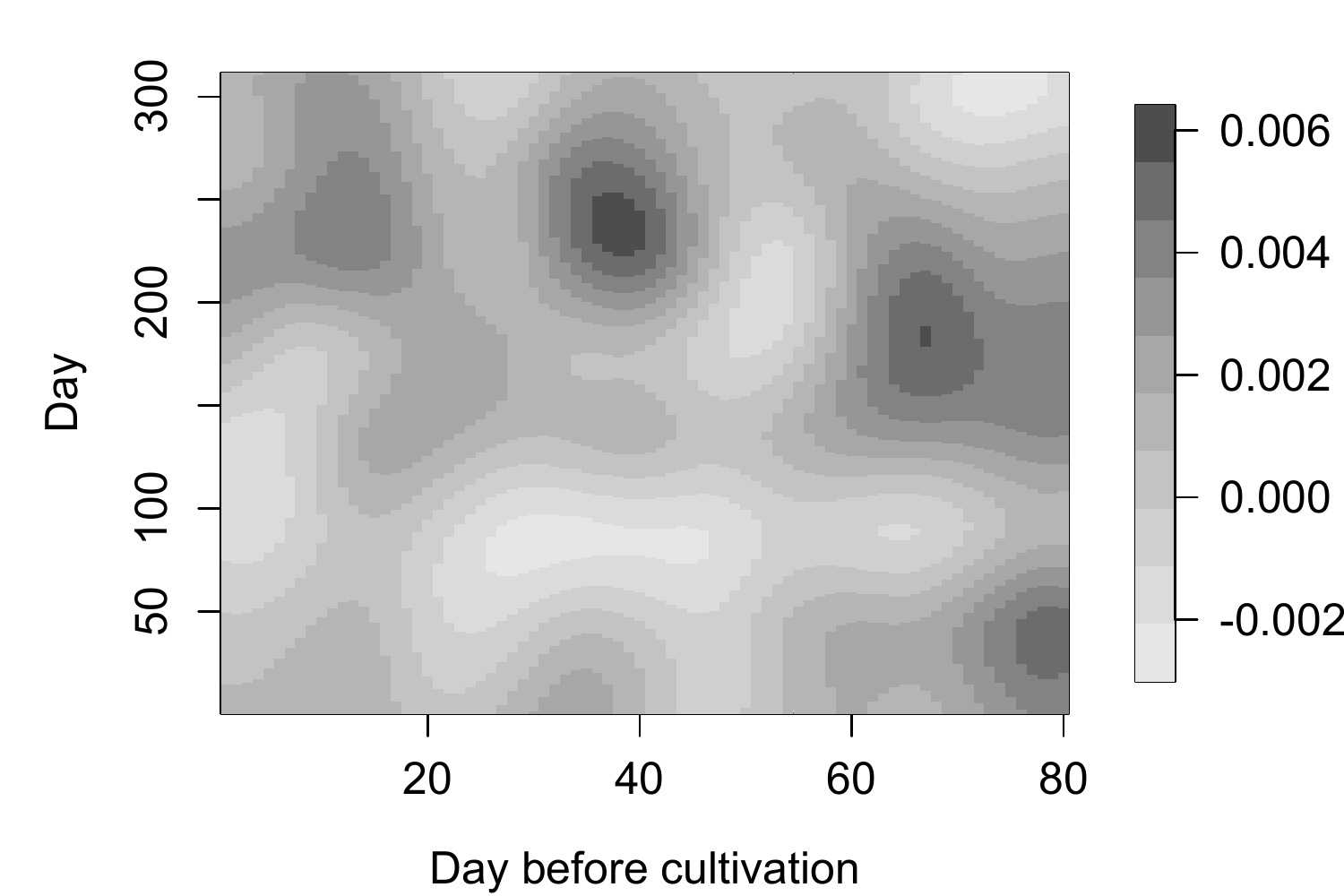}
		\includegraphics[width=0.48\hsize]{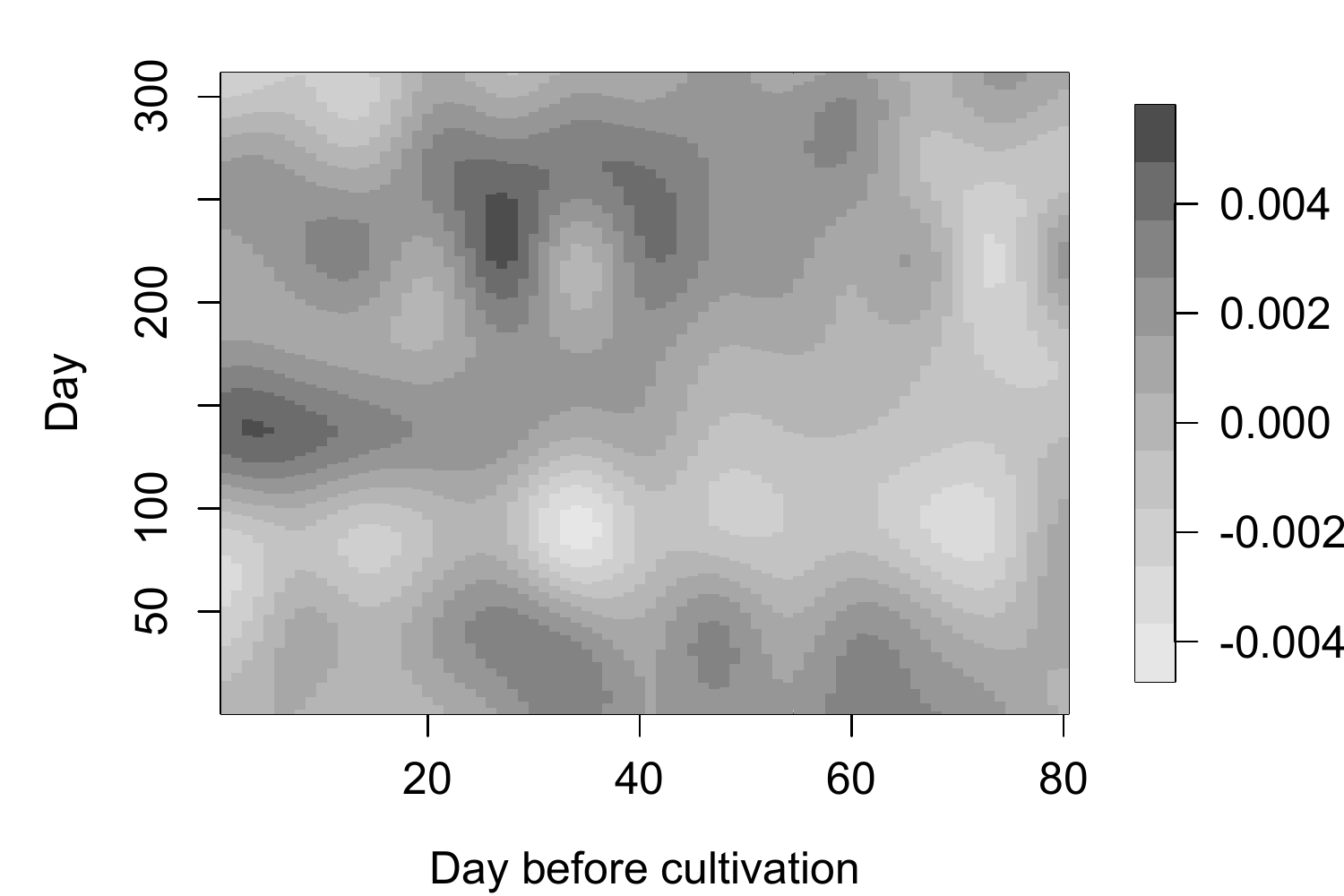}
	\end{center}
	\caption{Estimated coefficient surface $\hat{\beta}_j(s,t)$ for the temperature in the greenhouse (left) and the amount of solar radiation inside the greenhouse (right).} 
	\label{fig:surface}
\end{figure}

Figure \ref{fig:surface} shows estimated coefficient surface $\hat{\beta}_j(s,t)$ for the temperature in the greenhouse (Temp) and the amount of solar radiation inside the greenhouse (SunL).  
The horizontal axis represents the day before cultivation and the left side corresponds to the latest day.  
The vertical axis represents the days in calendar time during the growing season, with bottom representing early October and top representing middle of next July.
The coefficient of temperature indicates that
In late autumn and winter, the higher the temperature during the 80 days before fruiting, the lower the yield.
On the other hand, in spring and summer, the higher the temperature around 40 days before harvest, the higher the yield.
This may be related to the large increase in yield after spring.
A similar trend was obtained for the coefficient of solar radiation.
In autumn and winter, yields are higher when solar radiation is lower during the growing period, while in the spring and summer, yields are higher when solar radiation is higher 30 to 40 days before harvest.

\begin{table}[t]
	\centering
	\caption{List of environmental factors selected by SVCFLM.}
	\label{tab:env}
	\begin{tabular}{ll}
		\hline
		 Name & Description \\ 
		\hline
		Temp & Temperature inside the greenhouse \\ 
		Humi & Humidity inside the greenhouse \\ 
		Co2  & CO2 concentration inside the greenhouse \\
		Lux  & Illuminance outside the greenhouse \\
		SunL & Amount of solar radiation inside the greenhouse\\
		OutT & Temperature outside the greenhouse \\
		Wind & Wind speed outside the greenhouse \\
		WDir & Wind direction outside the greenhouse\\
		VPD & Vapor pressure deficit\\
		Opt-1 & Temperature of culture medium \\
		DPT & Dew point\\
		AT1YD & Average temperature one day before\\
		AT2YD & Average temperature two days before\\
		CV-K1 & Degree of ventilation opening 1\\
		CV-K2 & Degree of ventilation opening 2\\
		CV-C1 & Degree of curtain opening 1  \\
		CV-C2 & Degree of curtain opening 2\\
		CV-H1 & Heating unit operation 1\\
		CV-H2 & Heating unit operation 2\\
		CO2App & Amount of application of CO2\\
		Irri & Number of application of irrigation\\
		\hline
	\end{tabular}
\end{table}

\section{Concluding Remarks}
We have proposed SVCFLM, the method for variable selection in varying-coefficient functional linear models.  
The proposed method can simultaneously estimate parameters and select predictors in functional linear models, where the predictors are functions and the response are a scalar, and the response depends on an exogenous variable as well as functional predictors.  
The functional predictors are supposed to be expressed by truncated Kahhunen-Lo\'eve expansion, and then unknown parameters in CVCFLM are estimated by the adaptive elastic net penalty.  
We apply the blockwise descent algorithm to derive the estimators of these parameters.  
We also provided a model selection criterion for selecting appropriate values of tuning parameters.  
Simulation studies show the effectiveness of the proposed method in viewpoints of prediction accuracy and variable selection.  
We applied the proposed method to the analysis of crop yield data, and then selected the environmental factors that relates to the amount of crop yield.  

The relationship between the predictors and the response is linear in VCFLM, and therefore the flexibility of expressing this relationship is limited.  
One of the approaches for handling this issue is to extend the VCFLM to the additive model framework \cite{Ma2020}.  
However, such nonlinear models make it more difficult to obtain an interpretation about the relationships among variables.  
Future work reminds the extension of the proposed method to the additive model framework.

\section*{Acknowledgment}
We appreciate the Higashibaba farm for providing the data for cultivating tomatoes.  
This work was supported by JSPS KAKENHI Grant Number 19K11858.


\begin{thebibliography}{30}
	\expandafter\ifx\csname natexlab\endcsname\relax\def\natexlab#1{#1}\fi
	
	\bibitem[{Araki et~al.(2009)Araki, Konishi, Kawano, and
		Matsui}]{ArKoKa_etal2009a}
	Araki, Y., Konishi, S., Kawano, S., and Matsui, H. (2009),  {Functional
		regression modeling via regularized Gaussian basis expansions,}
	\textit{Ann. Inst. Stat. Math.} \bm{61}, 811--833.
	
	\bibitem[{Cardot and Sarda(2008)}]{CaSa2008}
	Cardot, H. and Sarda, P. (2008),  {Varying-coefficient functional linear
		regression models,} \textit{Comm. Statist. Theory Methods}, \bm{37}, 3186--3203.
	
	\bibitem[{Gertheiss et~al.(2013)Gertheiss, Maity, and Staicu}]{GeMaSt2013}
	Gertheiss, J., Maity, A., and Staicu, A.-M. (2013),  {Variable selection
		in generalized functional linear models,} \textit{Stat} \bm{2}, 86--101.
	
	\bibitem[{Hastie et~al.(2015)Hastie, Tibshirani, and Wainwright}]{HaTiWa2015}
	Hastie, T., Tibshirani, R., and Wainwright, M. (2015), \textit{Statistical
		Learning with Sparsity: The Lasso and Generalization}, Boca Raton: Chapman \&
	Hall/CRC.
	
	\bibitem[{Horv{\'a}th and Kokoszka(2012)}]{HoKo2012}
	Horv{\'a}th, L. and Kokoszka, P. (2012), \textit{Inference for functional data
		with applications}, New York: Springer.
	
	\bibitem[{James(2002)}]{Ja2002}
	James, G. (2002),  {Generalized linear models with functional
		predictors,} \textit{J. Roy. Statist. Soc. Ser. B} \bm{64},
	411--432.
	
	\bibitem[{Kayano et~al.(2016)Kayano, Matsui, Yamaguchi, Imoto, and
		Miyano}]{KaMaYa_etal2016}
	Kayano, M., Matsui, H., Yamaguchi, R., Imoto, S., and Miyano, S. (2016),
	 {Gene set differential analysis of time course expression profiles
		via sparse estimation in functional logistic model with application to
		timedependent biomarker detection,} \textit{Biostatistics} \bm{17}, 235--248.
	
	\bibitem[{Kokoszka and Reimherr(2017)}]{KoRe2017}
	Kokoszka, P. and Reimherr, M. (2017), \textit{Introduction to functional data
		analysis}, Boca Raton: CRC Press.
	
	\bibitem[{Matsui(2020)}]{Ma2020}
	Matsui, H. (2020),  {Varying-coefficient functional additive models,}
	\textit{arXiv preprint arXiv:2005.12641}.
	
	\bibitem[{Matsui and Konishi(2011)}]{MaKo2011}
	Matsui, H. and Konishi, S. (2011),  {Variable selection for functional
		regression models via the L1 regularization,} \textit{Comput.
		Statist. \& Data Anal.} \bm{55}, 3304--3310.
	
	\bibitem[{Matsui and Umezu(2020)}]{MaUm2020}
	Matsui, H. and Umezu, Y. (2020),  {{Variable selection in multivariate
			linear models for functional data via sparse regularization},}
	\textit{Jpn. J. Stat. Data Sci.} \bm{3}, 453--467.
	
	\bibitem[{Mclean et~al.(2014)Mclean, Hooker, Staicu, Scheipl, and
		Ruppert}]{McHoSt_etal2014}
	Mclean, M.~W., Hooker, G., Staicu, A.-m., Scheipl, F., and Ruppert, D.~R.
	(2014),  {{Functional Generalized Additive Models Functional
			Generalized Additive Models},} \textit{J. Comput. Graph. Statist.} \bm{23} 249--269.
	
	\bibitem[{Morris(2015)}]{Mo2015}
	Morris, J.~S. (2015),  {Functional regression,} \textit{Annu. Rev. Stat. Appl.} \bm{2}, 321--359.
	
	\bibitem[{M{\"u}ller and Stadtm{\"u}ller(2005)}]{MuSt2005}
	M{\"u}ller, H. and Stadtm{\"u}ller, U. (2005),  {Generalized functional
		linear models,} \textit{Ann. Statist.} \bm{33}, 774--805.
	
	\bibitem[{M{\"{u}}ller and Yao(2008)}]{MuYa2008}
	M{\"{u}}ller, H.-G. and Yao, F. (2008),  {{Functional Additive Models},}
	\textit{J. Am. Stat. Assoc.} \bm{103}, 1534--1544.
	
	\bibitem[{Ramsay and Dalzell(1991)}]{RaDa1991}
	Ramsay, J. and Dalzell, C. (1991),  {Some tools for functional data
		analysis,} \textit{J. Roy. Statist. Soc. Ser. B} \bm{53},
	539--572.
	
	\bibitem[{Ramsay and Silverman(2005)}]{RaSi2005}
	Ramsay, J. and Silverman, B. (2005), \textit{Functional data analysis (2nd ed.)}, New York: Springer.
	
	\bibitem[{Ravikumar et~al.(2009)Ravikumar, Lafferty, Liu, and
		Wasserman}]{RaLaLi_etal2009}
	Ravikumar, P., Lafferty, J., Liu, H., and Wasserman, L. (2009),  {Sparse
		additive models,} \textit{J. Roy. Statist. Soc. Ser. B}	\bm{71}, 1009--1030.
	
	\bibitem[{Reiss et~al.(2017)Reiss, Goldsmith, Shang, and
		Ogden}]{ReGoSh_etal2017}
	Reiss, P.~T., Goldsmith, J., Shang, H.~L., and Ogden, R.~T. (2017),
	 {{Methods for Scalar-on-Function Regression},} \textit{Int. Stat. Rev.} \bm{85}, 228--249.
	
	\bibitem[{Simon and Tibshirani(2012)}]{SiTi2012}
	Simon, N. and Tibshirani, R. (2012),  {Standardization and the Group
		Lasso Penalty,} \textit{Statist. Sinica} \bm{22}, 983--1001.
	
	\bibitem[{Tibshirani(1996)}]{Ti1996}
	Tibshirani, R. (1996),  {Regression shrinkage and selection via the
		lasso,} \textit{J. Roy. Statist. Soc. Ser. B} \bm{58}, 267--288.
	
	\bibitem[{Ullah and Finch(2013)}]{UlFi2013}
	Ullah, S. and Finch, C.~F. (2013),  {Applications of functional data
		analysis: A systematic review,} \textit{BMC Med. Res. Methodol.} \bm{13}, 43.
	
	\bibitem[{Wang et~al.(2007)Wang, Li, and Tsai}]{WaLiTs2007}
	Wang, H., Li, R., and Tsai, C. (2007),  {Tuning parameter selectors for
		the smoothly clipped absolute deviation method,} \textit{Biometrika} \bm{94},
	553--568.
	
	\bibitem[{Wu et~al.(2010)Wu, Fan, and M{\"u}ller}]{WuFaMu2010}
	Wu, Y., Fan, J., and M{\"u}ller, H. (2010),  {Varying-coefficient
		functional linear regression,} \textit{Bernoulli} \bm{16}, 730--758.
	
	\bibitem[{Yao et~al.(2005)Yao, M{\"u}ller, and Wang}]{YaMuWa2005b}
	Yao, F., M{\"u}ller, H., and Wang, J. (2005),  {Functional linear
		regression analysis for longitudinal data,} \textit{Ann. Statist.} \bm{33}, 2873--2903.
	
	\bibitem[{Yao and M\"uller(2010)}]{YaMu2010}
	Yao, F. and M\"uller, H.-G. (2010),  {Functional quadratic regression,}
	\textit{Biometrika} \bm{97}, 49--64.
	
	\bibitem[{Yuan and Lin(2006)}]{YuLi2006}
	Yuan, M. and Lin, Y. (2006),  {Model selection and estimation in
		regression with grouped variables,} \textit{J. Roy. Statist. Soc. Ser. B} \bm{68}, 49--67.
	
	\bibitem[{Zhang et~al.(2010)Zhang, Li, and Tsai}]{ZhLiTs2010}
	Zhang, Y., Li, R., and Tsai, C. (2010),  {Regularization parameter
		selections via generalized information criterion,} \textit{J. Am. Stat. Assoc.}
	\bm{105}, 312--323.
	
	\bibitem[{Zou and Hastie(2005)}]{ZoHa2005}
	Zou, H. and Hastie, T. (2005),  {Regularization and variable selection
		via the elastic net,} \textit{J. Roy. Statist. Soc. Ser. B} \bm{67}, 301--320.
	
	\bibitem[{Zou and Zhang(2009)}]{ZoZh2009}
	Zou, H. and Zhang, H. (2009),  {On the adaptive elastic-net with a
		diverging number of parameters,} \textit{Ann. Statist.}, 37,
	1733--1751.
	
\end{thebibliography}

\end{document}